# The competent Computational Thinking test (cCTt): Development and validation of an unplugged Computational Thinking test for upper primary school

Laila El-Hamamsy[1,2], María Zapata-Cáceres[3], Estefanía Martín Barroso[3], Francesco Mondada[1,2], Jessica Dehler Zufferey[2], Barbara Bruno[4]

**Abstract**
With the increasing importance of Computational Thinking (CT) at all levels of education, it is essential to have valid and reliable assessments. Currently, there is a lack of such assessments in upper primary school. That is why we present the development and validation of the competent CT test (cCTt), an unplugged CT test targeting 7-9 year-old students. In the first phase, 37 experts evaluated the validity of the cCTt through a survey and focus group. In the second phase, the test was administered to 1519 students. We employed Classical Test Theory, Item Response Theory, and Confirmatory Factor Analysis to assess the instruments' psychometric properties. The expert evaluation indicates that the cCTt shows good face, construct, and content validity. Furthermore, the psychometric analysis of the student data demonstrates adequate reliability, difficulty, and discriminability for the target age groups. Finally, shortened variants of the test are established through Confirmatory Factor Analysis. To conclude, the proposed cCTt is a valid and reliable instrument, for use by researchers and educators alike, which expands the portfolio of validated CT assessments across compulsory education. Future assessments looking at capturing CT in a more exhaustive manner might consider combining the cCTt with other forms of assessments.

## 1 Introduction and Related Work

### 1.1 The increasing position of CT in research and formal education

An international debate sparked with Wing (2006)'s article presenting Computational Thinking (CT) as a universally applicable attitude and skill set, as important as reading, writing and arithmetic. While a consensus has not been reached on the definition of CT, nor on where its boundaries lie, one prominent definition is the one by Brennan and Resnick (2012). Brennan and Resnick (2012) decompose CT into: i) computational concepts (the concepts most closely related to computer science and programming, i.e. "sequences, loops, parallelism, events, conditionals, operators, and data"), ii) computational practices (the strategies and practices required to be apply said concepts), and iii) computational perspectives ("the perspectives designers form about the world around them and about themselves"). While CT is traditionally considered to be the "thought processes that facilitate framing and solving problems using computers and other technologies" (Relkin and Bers 2021), more and more researchers argue that CT is not specific to "those interested in computer science and mathematics". These researchers consider that CT has a "multi-faceted theoretical nature" and can be considered more generally as an example of "models of thinking" (Li et al. 2020). Under this new light, CT is envisioned to have a broader role to play in education, from STEM-related disciplines (Li et al. 2020; Weintrop 2016; Peel et al. 2020), to languages (Rottenhofer et al. 2021), and transversal competences[1] such as "creative problem solving" (Grover et al. 2017; Chevalier et al. 2020). Some researchers thus consider CT to be one of the fundamental competences that every citizen must acquire

[1] MOBOTS Group, Ecole Polytechnique Fédérale de Lausanne, Switzerland
[2] LEARN – Center for Learning Sciences, Ecole Polytechnique Fédérale de Lausanne, Switzerland
[3] Universidad Rey Juan Carlos, Computer Science Department, Madrid, Spain
[4] Computer Human Interaction in Learning and Instruction (CHILI) Laboratory, Ecole Polytechnique Fédérale de Lausanne, Switzerland

**Corresponding author:**
Laila El-Hamamsy,
EPFL STI IEM SCI-STI-FMO1, ME B3 435 (Batiment ME), Station 9, CH-1015 Lausanne
+41 21 693 66 91
Email: laila.elhamamsy@epfl.ch

in the 21st century (Li et al. 2020). The growing importance of CT as a transversal competence has led, in the past decade, to an increase in the research around CT (Li et al. 2020; Ilic et al. 2018; Tang et al. 2020) as well as in the initiatives seeking to equip K-12 students with CT competences (Basu et al. 2020). Such initiatives operate through informal education (Weintrop et al. 2021), extra-curriculars, and formal education settings (Weintrop et al. 2021), some even starting at the level of kindergarten (Bocconi et al. 2016).

### 1.2 The need for CT assessments at all levels of education

With this "tremendous growth in curricula, learning environments, and innovations around CT education" (Weintrop et al. 2021), the design of tools to assess CT competences in a developmentally appropriate and reliable way throughout compulsory education becomes crucial (Hsu et al. 2018). Indeed, "CT assessment is important [to document] learning progress, [measure] lesson effectiveness, [assist] in curriculum development and [help] identify students in need of greater assistance or enrichment" (Relkin and Bers 2021). As CT is also becoming popular among K-9 educators (Chen et al. 2017; Mannila et al. 2014), it is also paramount to develop assessments for that age range. Indeed, as mentioned by Relkin and Bers (2021), "one of the greatest challenges to integrating CT into early elementary school education has been a lack of validated, developmentally appropriate assessments to measure young students' CT skills in classroom and online settings" (Lockwood and Mooney 2017; Román-González et al. 2019). Assessment tools should therefore be adapted for use, not only by researchers looking to design CT learning experiences (Weintrop et al. 2021) and investigate how best to foster CT competences (Chevalier et al. 2020), but also by teachers aiming to ensure that their students are acquiring the desired competences, and this starting from kindergarten onward (Zapata-Cáceres et al. 2020). Provided the pressing need to clarify the question about how best to assess CT competences (Tang et al. 2020; Lockwood and Mooney 2017; Hsu et al. 2018), it is not surprising to find that CT assessment "is at the forefront of CT research [and] gathering the greatest interest of researchers" (Tikva and Tambouris 2021). However, as stated by Zapata-Cáceres et al. (2020) and Román-González et al. (2019), most efforts to develop assessments for CT have focused on secondary school and tertiary education.

### 1.3 The importance of validated and reliable instruments which do not conflate with programming abilities

Developing CT assessments must consider, in addition to the developmental appropriateness for the target age group, the different assessment formats which exist, their use cases, and their scalability. From a design perspective, four main formats have been used to assess CT (Tang et al. 2020): traditional tests (often used in combination with other assessment methods), portfolios (to "[situate] CT assessment in a real-world context and further allows teachers and researchers to provide formative feedback"), interviews ("to support or elaborate on the results of traditional or portfolio assessment by specifying students' thinking processes") and surveys (to assess dispositions and attitudes towards CT). The most common approach seems to consist in the use of portfolios to "analys[e] projects performed by students in specific programming environments" (Tang et al. 2020). Unfortunately, assessment tools following this approach carry the risk of "conflating CT with coding abilities" (Relkin and Bers 2021), which may limit their use in i) pre-post test designs (Chen et al. 2017), ii) when validating new learning environments, and iii) in cases where Computer Science (CS) unplugged activities (Bell and Vahrenhold 2018) are employed to develop students' CT competences[2]. Because of the transversal nature of computational thinking and the "variety of methods by which CT is taught and contexts in which students learn CT" (Weintrop et al. 2021), researchers advocate the development of CT tests that go beyond self-reporting, and are more general (Tikva and Tambouris 2021). Such assessments should be agnostic of the specific content of the study and the programming environments. The past few years have thus seen a rise in age appropriate unplugged CT assessments targeting CT skills[3] which

i) can be administered without employing screens (referring to the definition of Unplugged provided by Bell and Vahrenhold 2018) and can thus be easily deployed in various settings and at a large scale,
ii) do not require any prior knowledge pertaining to programming or coding (including that of a specific programming language) and are therefore adapted for use in pre-post test experimental designs,
iii) put a strong emphasis on the reliability and validity said instruments (Román-González et al. 2017, 2018, 2019; Zapata-Cáceres et al. 2020; Relkin et al. 2020; Chen et al. 2017; Wiebe et al. 2019), something which has been identified so far as lacking in the CT assessment literature (Tang et al. 2020).

*1.4  The lack of existing validated and reliable CT assessments spanning compulsory education*

Considering the importance of having unplugged CT assessments which are i) agnostic of programming skills and ii) have undergone psychometric analyses for validity and reliability, we find ourselves lacking full coverage from kindergarten to upper secondary school (see Table 1).

Starting from the lower end of the spectrum, the TechCheck-K was recently developed by Relkin and Bers (2021) to assess CT at the level of kindergarten, considering the requirements of that age group in terms of cognitive, literacy and motor development.

Two assessment tools exist for lower primary school: the TechCheck (Relkin et al. 2020) and the Beginner's CT test (BCTt, Zapata-Cáceres et al. 2020). The TechCheck (Relkin et al. 2020) was developed for grades 1-2 and proved reliable through classical test theory and item response theory and valid in comparison with the original TACTIC-KIBO instrument (Relkin et al. 2020), thus speaking to the instruments' convergent validity. The BCTt draws inspiration from the CT test (CTt, Román-González et al. 2017, 2019), specifically adapting the original CTt in terms of format and content to take into account students' limited reading and understanding skills (Tikva and Tambouris 2021; Zhang and Nouri 2019) and ensure the use of "developmentally appropriate language and tasks to assure that factors such as literacy and fine motor skills are not limiting" (Relkin et al. 2021). The instrument, which follows a multiple choice format, was validated with both experts and primary school students (ages 5-10) without prior coding experience, but showed a ceiling effect for students aged 7-10, with the developmental appropriateness in regards to the length of the test (45 minutes) for lower primary being put into question by (Relkin et al. 2021). Indeed, provided the objective of having developmentally appropriate instruments, which can also be used in a diverse range of settings, including researchers in pre-post intervention study designs, and practitioners evaluating the impact of educational reforms (notably digital and computing education), length of administration also becomes a major factor of adoption.

While certain researchers have looked into developing assessments for upper primary school (Gane et al. 2021; Parker et al. 2021), they suffer from numerous limitations. The Bebras challenge (Román-González et al. 2017) for example is an international competition for students throughout compulsory school. While it is sometimes used to assess CT skills, it has undergone limited psychometric validation (Hubwieser and Mühling 2014; Bellettini et al. 2015). At the same time, the assessment by Gane et al. (2021) requires manual grading and multiple annotators, limiting the test's scalability. Finally, the assessment by Parker et al. (2021) which relies on a combination of block-based and Bebras-style questions, has been piloted with just 57 4th graders.

Finally, we find the CTt by Román-González et al. (2017, 2018, 2019), a multiple choice test designed and adapted for secondary school (ages 10-16). The CTt has undergone several stages of validation (including reliability, criterion validity in relation to other cognitive tests Román-González et al. 2017, predictive validity Román-González et al. 2018, and convergent validity Román-González et al. 2019).

There thus appears to be a gap in validated unplugged assessments for students in upper primary school (ages 7-9, Román-González et al. 2017).

*1.5  Research questions*

Therefore, to expand the portfolio of validated CT tests across compulsory education, specifically filling the gap in upper primary school, in this article we present the competent CT test (cCTt), the steps undertaken to develop it on the basis of the BCT test, and its psychometric validation with 37 experts and 1519 students. In the following, we specifically consider the following research questions:

- RQ1: Is the cCTt a valid measure of CT skills for students in third and fourth grade (aged 7 to 9)?
- RQ2: Is the cCTt a reliable measure of CT skills for students in third and fourth grade (aged 7 to 9)?
- RQ3: Is the length of the cCTt developmentally appropriate for students in third and fourth grade? If not, can it be shortened?

**Table 1.** Synthesis of decontextualised unplugged CT assessments and corresponding validation processes.

| Test | Format | Target age group | Underlying CT definition | Validation process | Sample | Validity established for |
|---|---|---|---|---|---|---|
| TechCheck-K (Relkin and Bers 2021) | 15 item MCQ | kindergarten (5-6 year old students) | Algorithms, Modularity, Design Process, Debugging, Control Structures, Hardware/Software | Classical test theory | 89 kindergarten students without coding experience | Full sample |
| TechCheck (Relkin et al. 2020) | 15 item MCQ | 1st and 2nd grade (6-9 year old students) | | Expert validation (face validity), psychometric analysis (reliability and validity through Classical Test Theory - IRT) and convergent validation Theory - IRT) and convergent validation with the TACTIC-KIBO | 768 5-9 year old students participating in a robotics coding curriculum | Full sample |
| Beginner's CT test (Zapata-Cáceres et al. 2020) | 25 item MCQ | Primary school (5-12 year old students) | Computational concepts, practices, perspectives (Brennan and Resnick 2012) | Expert validation, and psychometric analysis (reliability and validity - Classical Test Theory) | 299 primary school students from grades 1-6 | 5-7 year old students |
| Bebras challenge (Hubwieser and Mühling 2014) | 15 item MCQ (3 sets of 5 tasks for 6 age groups with easy, medium & difficult questions) | Secondary school (10-16 year old students) | Problem solving, algorithm design, pattern recognition, pattern generalisation and abstraction (see bebras.org) | Psychometric analysis (IRT) to identify constructs | 55088 10-16 year old students | |
| Bebras challenge (Bellettini et al. 2015) | | | | Psychometric analysis (IRT) to determine the difficulty of the questions | 2736 students aged 11-18 | |
| [unnamed test] (Gane et al. 2021) | 6 instruments ( 10 items each) with assessment rubrics | Grades 3-4 (age not specified) | Sequence, repetition, conditionals, decomposition, variables, and debugging | Inferential (Classical Test Theory and IRT) and cognitive (design methodology) validation | 144 grade 3 and 4 primary school students following an integrated instruction | 2 instruments for students in grades 3-4 |
| ACES (Parker et al. 2021) | 10 item MCQ including block based and Bebras style questions | Grades 3-5 (age not specified) | Sequences and loops | Cognitive interviews and psychometric analysis through classical test theory | 57 4th grade students who completed a CT curriculum | Full sample |
| [unnamed test] (Chen et al. 2017) | 23 items (15 MCQ, 8 open ended with assessment rubrics) | Grade 5 (age not specified) | CT as "thinking in a way that can be represented and processed by machines to formulate and solve problems" | Psychometric analysis through (IRT, construct validity) and demonstration of student learning (pre-post test) | 121 5th grade students following a robotics curriculum | Full sample |
| CT test (González 2015) | 28 item MCQ with Scratch style questions | Secondary school (10-16 year old students) | Computational concepts, practices, perspectives (Brennan and Resnick 2012) | Expert validation | | |
| CT test (Román-González et al. 2017) | | | | Psychometric analysis (Classical Test Theory) and criterion validity (comparison with spatial, reasoning, and problem solving abilities) | 1,251 middle school students (10-16 years old) | Full sample |
| CT test (Román-González et al. 2018) | | | | Predictive validity | 314 middle school students (12-14 years old) | Full sample |

## 2 Methodology

The objective of the study was to design a CT assessment adapted for students in upper primary school, with adequate psychometric properties: the competent CT test (cCTt). In section 2.1 we describe the format of the cCTt and its development, before presenting its validation in section 2.2. Validation was done in two stages. First, the cCTt underwent an expert evaluation (see section 2.2.1). Subsequent adjustments were made based on their recommendations before administering the cCTt to over 1517 students to evaluate the psychometric properties of the test (see section 2.2.2).

### *2.1 The competent Computational Thinking Test (cCTt)*

The cCTt was developed by adapting the BCTt in terms of format and content to the target age group, as was done by researchers to develop the TechCheck-K from the TechCheck (Relkin et al. 2021) and to develop the BCT test from the CT test (Zapata-Cáceres et al. 2020). Both the BCTt and the cCTt are unplugged (i.e., paper-based) multiple choice exams composed of 25 questions of progressive difficulty which employ questions in two formats (see Fig. 1). The majority of questions (21 out of 25) use a 3x3 or a 4x4 grid that a chick must navigate to reach a hen, possibly satisfying side-goals as well, such as picking up a flower or avoiding a cat. The remaining questions are canvas-type questions where students have to replicate a drawing pattern. In both formats, each question presents itself with four possible answers from which the students must choose from. As shown in Table 2, both tests address CT concepts, as defined by Brennan and Resnick (2012), by successively assessing notions of sequences, simple loops, nested loops, conditionals (if-then and if-then-else), and while statements, but with varying number of questions per concept. While designing the cCTt, adaptations in terms of *content* aimed at rendering the test more complex by

i) mainly employing 4x4 grids,
ii) removing questions of low difficulty (i.e. which exhibited notable ceiling effects in Zapata-Cáceres et al. 2020, in particular questions on 3x3 grids and questions involving sequences and simple loops)
iii) adding more questions related to complex concepts (e.g. while statements)
iv) creating a new subset of questions which looks to determine whether students had assimilated the range of concepts addressed in the test (referred to as "combinations")
v) altering the disposition of objects (starting point, targets and or obstacles) on the grid, and/or the selection of responses students could choose from, so that identifying the correct response requires more reflection on the students' part.

Adaptations in terms of *format* primarily concerned the while statements: to convey the notion of repetition, and to make the statement more clearly distinct from the simple sequences, the symbols were adapted as shown in Fig. 2.

The individual questions of the cCTt are described in Table 3 and the full subset of questions is provided in appendix A.

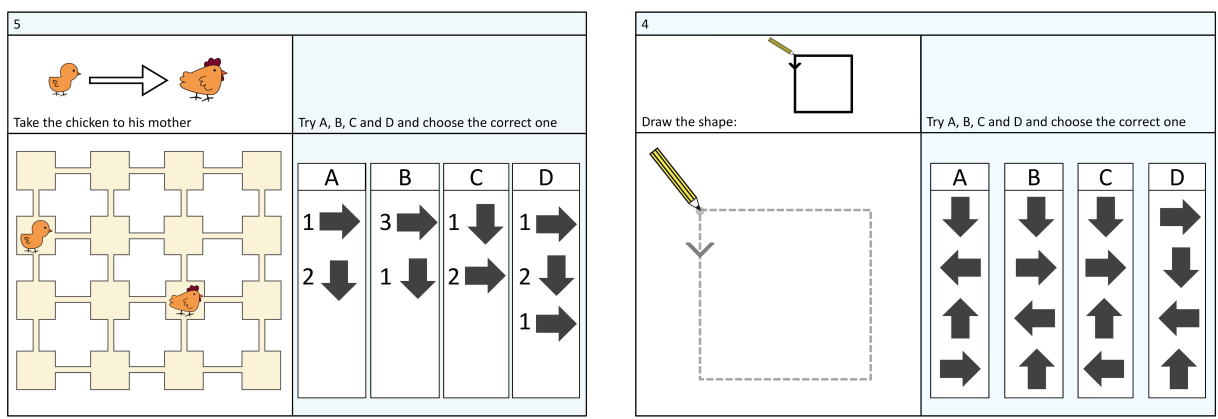

**Figure 1.** The two main question formats of the cCTt: grid (left) and canvas (right).

**Table 2.** Comparison between the BCTt and the cCTt in terms of question concepts and question types.

| Blocks | BCTt | | | | cCTt | | | |
|---|---|---|---|---|---|---|---|---|
| | Grid (3x3) | Grid (4x4) | Canvas | Total | Grid (3x3) | Grid (4x4) | Canvas | Total |
| Sequences | 3 | 1 | 2 | 6 | 1 | 1 | 2 | 4 |
| Simple loops | 3 | 2 | 0 | 5 | 0 | 4 | 0 | 4 |
| Complex loops | 0 | 5 | 2 | 7 | 0 | 5 | 2 | 7 |
| Conditional statements | 1 | 3 | 0 | 4 | 1 | 3 | 0 | 4 |
| While statements | 1 | 2 | 0 | 3 | 1 | 3 | 0 | 4 |
| Combinations | 0 | 0 | 0 | 0 | 0 | 2 | 0 | 2 |
| Total | 8 | 13 | 4 | 25 | 3 | 18 | 4 | 25 |

**Table 3.** Description of the cCTt, as administered to the students for validation. Note that this format includes error profiles associated to each response (see section 2.2.1 and 3.1.1), with similar distributions of profiles across options A, B, C and D, as suggested by the experts in the expert evaluation. Q: question, O : obstacles, P : pickups.

| Block (CT concept) | Q | O | P | Format | Profile for answer | | | | Profile 1 | Profile 2 | Profile 3 | Profile 4 |
|---|---|---|---|---|---|---|---|---|---|---|---|---|
| | | | | | A | B | C | D | | | | |
| Sequences | 1 | x | x | Grid (3x3) | 3 | 4 | 2 | 1 | Incorrect interpretation of the displacements | Correct displacements but lacking objectives | Partial error with the objectives | Correct |
| | 2 | x | - | Canvas | 2 | 3 | 1 | 4 | | | | |
| | 3 | - | x | Grid (4x4) | 4 | 1 | 3 | 2 | | | | |
| | 4 | - | - | Canvas | 1 | 2 | 4 | 3 | | | | |
| Simple loops | 5 | | | Grid (4x4) | 1 | 3 | 4 | 2 | Ignoring the distances / numbers | Incorrect target | Incomplete objectives | Correct |
| | 6 | | | Grid (4x4) | 4 | 2 | 3 | 1 | | | | |
| | 7 | x | x | Grid (4x4) | 3 | 1 | 2 | 4 | | | | |
| | 8 | x | x | Grid (4x4) | 2 | 4 | 1 | 3 | | | | |
| Complex loops | 9 | | | Grid (4x4) | 3 | 4 | 2 | 1 | Ignoring the loops | Incorrect target | Incomplete objectives | Correct |
| | 10 | | - | Canvas | 1 | 2 | 4 | 3 | | | | |
| | 11 | | | Grid (4x4) | 4 | 2 | 1 | 3 | | | | |
| | 12 | x | x | Grid (4x4) | 4 | 3 | 2 | 1 | | | | |
| | 13 | x | x | Grid (4x4) | 2 | 1 | 3 | 4 | | | | |
| | 14 | - | - | Canvas | 1 | 3 | 4 | 2 | | | | |
| | 15 | x | x | Grid (4x4) | 2 | 1 | 3 | 4 | | | | |
| Conditional statements | 16 | | | Grid (3x3) | 2 | 3 | 1 | 4 | Ignoring the conditional statements | Incorrect target, misinterpretation of the statements | Interpretation as a while | Correct |
| | 17 | | | Grid (4x4) | 1 | 2 | 4 | 3 | | | | |
| | 18 | | | Grid (4x4) | 3 | 4 | 2 | 1 | | | | |
| | 19 | | | Grid (4x4) | 4 | 1 | 3 | 2 | | | | |
| While statements | 20 | | | Grid (3x3) | 1 | 2 | 4 | 3 | Ignoring the while statements | Starting point misinterpretation | Interpretation as a go to | Correct |
| | 21 | | | Grid (4x4) | 3 | 4 | 2 | 1 | | | | |
| | 22 | | | Grid (4x4) | 2 | 3 | 1 | 4 | | | | |
| | 23 | | | Grid (4x4) | 4 | 1 | 3 | 2 | | | | |
| Combinations | 24 | | | Grid (4x4) | 1 | 2 | 3 | 4 | Ignoring the symbols | Misinterpreting the while | Misinterpreting distances & loops | Correct |
| | 25 | | | Grid (4x4) | 3 | 1 | 4 | 2 | | | | |

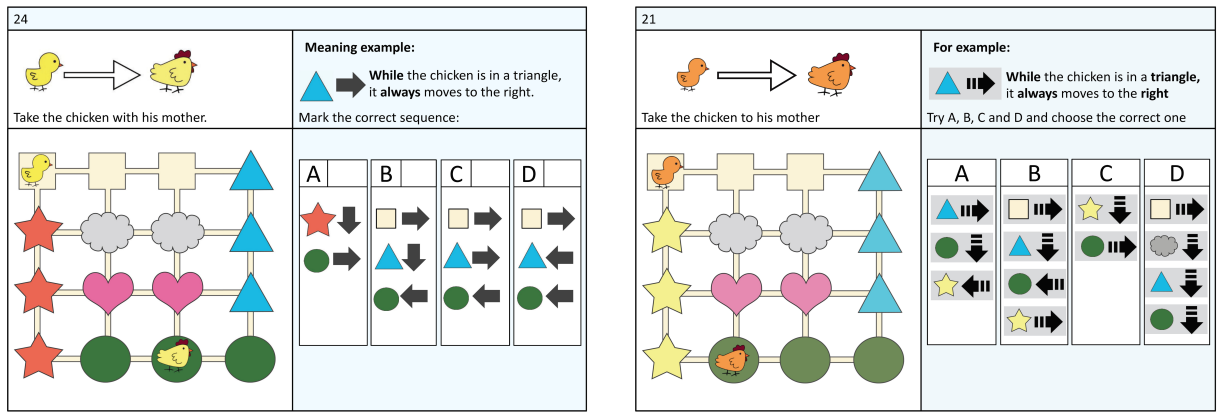

**Figure 2.** Example of a format and content adaptation (Q24 in the BCTt on the left, which became Q21 in the cCTt on the right), including the recommendations made by the experts.

## *2.2 Validation of the cCTt*

The validation of instruments generally falls under the field of psychometrics, which we briefly introduce here before detailing the approach undertaken for the cCTt.

Psychometric theories are part of a field which seeks to understand the structure of intelligence and "portrays intelligence as a composite of abilities measured by mental tests"[4]. Psychometric theories help evaluate the quality of assessments through two main properties: validity and reliability. Reliability is "the ability to reproduce a result consistently in time and space". Validity on the other hand "refers to the property of an instrument to measure exactly what it proposes" (Souza et al. 2017) and is typically presented under four forms (Taherdoost 2016):

- Construct validity, or the "extent to which there is evidence consistent with the assumption of a construct of concern being manifested in subjects' observed performance of the instrument" (Raykov and Marcoulides 2011). Simply put, does our test measure the skills/abilities it intends to measure?
- Content validity, or the "degree to which test components represent adequately a performance domain or construct of interest" (Raykov and Marcoulides 2011), i.e. "whether the particular items in the test adequately represent the domain of possible items one could construct" (Raykov and Marcoulides 2011). Simply put, is the test fully representative of the content it aims to measure?
- Criterion validity, or how closely the results of the test correspond to the results of a different test.
- Face validity, or the extent to which the content of the test appears to be suitable to its aims.

Both validity and reliability must be considered to adequately validate an assessment instrument. As such, we employ expert evaluation to look into the face, construct and content validity of the test (see section 2.2.1) and analyse student performance to establish both construct validity and reliability of the cCTt (see section 2.2.2). Criterion validity, generally established by comparing the assessment with other validated instruments (Román-González et al. 2017, 2019), is lacking in the present study.

*2.2.1 Expert evaluation* is an approach that has been used by many researchers to validate CT assessments (Djambong et al. 2018; Zapata-Cáceres et al. 2020; Relkin et al. 2021). That is why, prior to administering the test to the students, a panel of experts was organised to i) "evaluate if the test [content corresponds] to the intended constructs" (Tang et al. 2020, construct validity), ii) if the test is an adequate measurement of CT competence (content validity) and iii) whether the test appears adequate for students in upper primary school (face validity). In total, 37 experts of diverse backgrounds participated in the evaluation of the cCTt (see Table 4). The experts were recruited from:

1. The panel of experts having participated in the validation of the BCTt (Zapata-Cáceres et al. 2020).
2. Researchers in education and computer science working on the question of fostering and/or assessing CT.
3. Practitioners involved in the local digital education reform to gain additional insight into the developmental appropriateness of the instrument. These included a) experts in assessment from the department of education, b) professors from the university of teacher education and c) teachers having been trained to introduce digital education (including computer science and CT) and presently teaching said concepts to students in the target age group.

The expert evaluation was conducted in two stages. First, the experts were asked to respond to the survey described in Table 5, with the possibility to leave comments and provide suggestions for improvements after each block of questions. In a second stage, the experts were invited to a short presentation of the development of the test and objectives, which was followed by an unstructured focus group to discuss the experts' impressions of the test and their suggestions for improvements.

**Table 4.** Experts' profiles. Note that the experts had the possibility to select multiple occupations. Two early childhood education teachers were part of the experts who have been teaching CT to their students for years and participated in the evaluation of the BCTt (Zapata-Cáceres et al. 2020), while the other one is a former primary school teacher who is presently working as a teacher trainer and doing a PhD in a CT-related field.

| Occupation and expertise | Number of experts |
| --- | --- |
| Early Childhood Education Teacher | 3 |
| Elementary Education Teacher | 5 |
| High School Teacher | 1 |
| Teacher in the field of computer science or information technology | 4 |
| Teacher trainer | 5 |
| Background in computer science or information technology (or related) | 4 |
| PhD Student in education | 5 |
| Researcher in education | 10 |
| Teacher / Professor at a University | 13 |
| Professor at a Teacher Education University | 1 |
| Undisclosed | 7 |
| Age | Number of experts |
| ≤ 30 | 2 |
| Between 31 and 45 | 21 |
| Between 46 and 60 | 8 |
| Undisclosed | 6 |
| Gender | Number of experts |
| Female | 15 |
| Male | 16 |
| Undisclosed | 6 |
| Expertise teaching computer science, programming and / or computational thinking | 5-point Likert scale |
| (from 1 = no knowledge to 5 = expert) | $\mu = 3.7$ |

**Table 5.** Questions used in the expert evaluation survey.

| | Question | Validity Type | Format |
|---|---|---|---|
| Evaluation of the individual questions of the cCTt | Please select the correct response | | Multiple choice |
| | Indicate in your opinion the level of difficulty of the question for students aged 7-10 | Face | 7 Point Likert |
| Evaluation of the individual blocks of the cCTt | In your opinion, the questions you just did are adapted to measure students' knowledge of the [concept] | Construct | 7-Point Likert |
| | In your opinion, the questions you just did are relevant to evaluation computational thinking skills [5] | Content | 7-Point Likert |
| | Have clear illustrations and instructions | Face | 7-Point Likert |
| | We are open to any suggestions you may have to help improve the quality of the test | Face, construct and content | Open-ended |
| Evaluation of the cCTt | Do you think the number of questions is adequate? (yes, too much, not enough) | Face | Multiple choice |
| | Indicate, in your opinion, the level of adequacy of the test for students aged 7-10 | Face | 5-Point Likert |
| Demographics | Indicate your knowledge in terms of teaching computer science and or programming, and computational thinking | | 5-Point Likert |
| | Age, gender, occupation | | Checkboxes |

### 2.2.2 Psychometric reliability and construct validation from student data

*Administration* of the cCTt was done after implementing the adjustments suggested by the experts (see section 3.1.1). Using a paper-based format, generated with the Auto-Multiple-Choice software[6], 1519 students (77 classes) in grade 3 and grade 4 (ages 7-9) did the cCTt in February 2021 (see Table 6). The students belong to 7 establishments in the Canton Vaud in Switzerland that were selected to be representative of the demographics of the region (El-Hamamsy et al. 2021b). Due to COVID-19 restrictions the researchers could not be present to administer the tests, which is why a detailed protocol was provided to teachers. In particular, teachers were asked to demonstrate an explanatory example (see appendix B) on the board for each computational concept, and adhere to the recommended time frame (i.e. 40 minutes and 1.5 hours maximum as in the BCTts). The data was then scanned and integrated into a spreadsheet using the Auto-Multiple-Choice software. Additional data pre-processing for the validity and reliability analyses included replacing the lack of responses, "I don't know" and multiple checked-boxes with a score of 0.

*Construct validity* was assessed through factor analysis, a technique which groups "observed variables [(here, the questions of the test)] into latent [variables, (here, the associated concept)] based on commonalities within the data" (Atkinson et al. 2011). Two main approaches exist for factorial analyses (exploratory and confirmatory). Confirmatory Factor Analysis (CFA) is used when there is an assumption about the underlying structure of the data and to "confirm the structural model of an instrument" (de Souza et al. 2019), while exploratory factor analysis is usually used to explore the dimensionality of the data at hand. In our case, the objective being to evaluate whether the blocks of questions presented in Table 3 constitute coherent groups of questions (i.e. factors, or latent variables which cannot be directly observed - e.g. intelligence, motivation, happiness - but are inferred from others), we employed CFA. When conducting CFA one must employ multiple fit indices as they provide "a more holistic view of goodness of fit, accounting for sample size, model complexity, and other considerations relevant to the particular study" (Alavi et al. 2020). Two types of fit indices exist and must be employed in parallel : i) global model fit indices assess "how far a hypothesized model is from a perfect model" (Xia and Yang 2019) (such as the chi-square $\chi^2$ statistic, the root mean square error of approximation or RMSEA, and standardised root mean square residual or SRMR), while ii) local or incremental fit indices "compare the fit of a hypothesized model with that of a baseline model (i.e., a model with the worst fit)" (Xia and Yang 2019) (such as the comparative fit index CFI and the Tucker-Lewis index TLI). Cutoffs frequently employed for these metrics are (Xia and Yang 2019; Hu and Bentler 1999) CFI and TLI > 0.95, RMSEA < 0.06, SRMR < 0.08 $p_{\chi^2}$ > 0.05. It is important to point out that these thresholds are conventional and were established in a specific context, thus meaning that they should be considered as indicative of levels of miss-specification, rather than absolute judges of bad model fit (Xia and Yang 2019). Additionally, the $\chi^2$ statistic is sensitive to sample size, with larger samples decreasing the p-value (Alavi et al. 2020; Prudon 2015), and should thus not be considered as "the final word in assessing fit" (West, Taylor, & Wu, 2012, p. 211)" (Alavi et al. 2020). Some authors therefore suggest employing the ratio between the $\chi^2$ statistic and the degrees of freedom with a cutoff at $\chi^2 \leq 3$ (Kyriazos 2018). Finally, as the input data is binary (with a score of 0 or 1 per question), the CFA analysis is conducted using an estimator which is adapted to non normal data and employs diagonally weighted least squared to estimate the model parameters and tetrachoric correlations (Schweizer et al. 2015; Rosseel 2020).

**Table 6.** Student demographics.

|  | Grade 3 | Grade 4 | Grade 3-4 (mixed classrooms) |
|---|---|---|---|
| Female | 337 | 372 | 28 |
| Male | 386 | 385 | 31 |

*Reliability* was evaluated through two means: Classical Test Theory and Item Response Theory (IRT), as recommended in the context of instrument validation (Relkin et al. 2020; Cappelleri et al. 2014; Embretson and Reise 2000). Classical Test Theory "comprises a set of principles that allow us to determine how successful our proxy indicators are at estimating the unobservable variables of interest" (DeVellis 2006) and focuses on test scores (Hambleton and Jones 1993). Main metrics include i) difficulty (proportion of students responding correctly), ii) reliability (proportion of the item's variation that was shared with the true score, often computed using Cronbach's alpha when considering scale reliability) and iii) discrimination (i.e. to what extent the question helps distinguish between the top performers and the low performers, estimated using the Point-biserial correlation). Unfortunately, Classical Test Theory suffers from several limitations. As its focus is at the test level, the observed scores and true scores are test-dependent, and sample-dependent (Hambleton and Jones 1993). In other words, "different samples with different variances will not yield equivalent data or data that can easily be compared across samples" (DeVellis 2006). While Classical Test Theory can be used to compare groups against one another, this can also put into question the reliability of the test. Moreover, "a score value on one item should mean the same thing as the same score value on another item of the same scale" (DeVellis 2006), which is not necessarily true when we consider assessments that have questions of increasing difficulty. That is why researchers have advocated the use of ability scores which are test independent. Item Response Theory (IRT) addresses this limitation, as it helps estimate the ability of an examinee, which is test independent. IRT is based on latent trait theory and assumes that there is an underlying student ability which leads to consistent performance, i.e. that the probability of a student getting a given item correct is a function of said students' ability. IRT thus operates at the item level and estimates parameters for the population (and not just the sample), but is based on high assumptions which are often hard to meet (Hambleton and Jones 1993).

The data analysis was conducted in R (version 3.6.0, R Core Team 2019) using R Studio (version 1.2) with the lavaan (version 0.6-7, Rosseel 2012), CTT (version 2.3.3, Sheng 2019), psych (version 2.1.3, Revelle 2021) and ltm (version 1.1.1, Rizopoulos 2006) packages.

### 2.3 Shortening the cCTt

A further objective of our work is to determine whether the length of the test if developmentally appropriate through the expert evaluation and by gathering feedback from teachers in the field having participated in the test's administration. Provided their responses, if inadequate, the objective is to propose several versions of the cCT test which can be selected according to researchers' and/or practitioners' needs. To that effect, Confirmatory Factor Analysis will be employed once more, this time not to validate the underlying structure of the test, but to identify questions exhibiting high correlations with other factors or other questions. Indeed, CFA fit indices tend to improve with parsimonous models, i.e. which are less complex (Alavi et al. 2020).

## 3 Results

### 3.1 RQ1 - Validity of the cCTt

As anticipated, we establish the validity of the cCTt through two means: expert validation, reported in section 3.1.1, and Confirmatory Factor Analysis, detailed in in section 3.1.2. Considering the validity measures, the experts appear in agreement on the face and content validity of the test, and both the expert validation and the factor analysis point to adequate construct validity. While content validity is evaluated positively, the experts highlight that the test is best suited for computational concepts and does not measure the full range of competences related to CT, notably computational practices (i.e. thought processes) and computational perspectives (i.e. perception). This is indeed an identified limitations of summative assessments (Román-González et al. 2019), which, if one desires a more exhaustive CT assessment, may be addressed by combining multiple assessment methods (Grover et al. 2015; Román-González et al. 2019).

### 3.1.1 Expert validation

*Face Validity* of the cCTt was evaluated based on several criteria. The evaluation of each individual question's difficulty points to a progressive increase in difficulty as intended, with an average overall difficulty of −0.4±1.0% (between "neither easy nor difficult" and "somewhat difficult" on a scale of -3 to +3). The experts, unaware that there was a protocol with examples that was provided to teachers at the time of the survey administration, believed that the symbolism would be difficult to understand without prior explanations, notably in the case of the for loops, the if-else statements, while statements and their combinations. This accounts for the more negative ratings obtained in terms of illustrations for those blocks of questions, and confirms the importance of providing examples to the students beforehand to grasp the key mechanics of the test. As for the test length, 52% of respondents believed it was adequate, 41% that it was too long, and 7% that it was not enough. All in all, 63% believed the test was adequate to measure upper primary school students' CT skills, 26% were neutral, and just 11% were in disagreement.

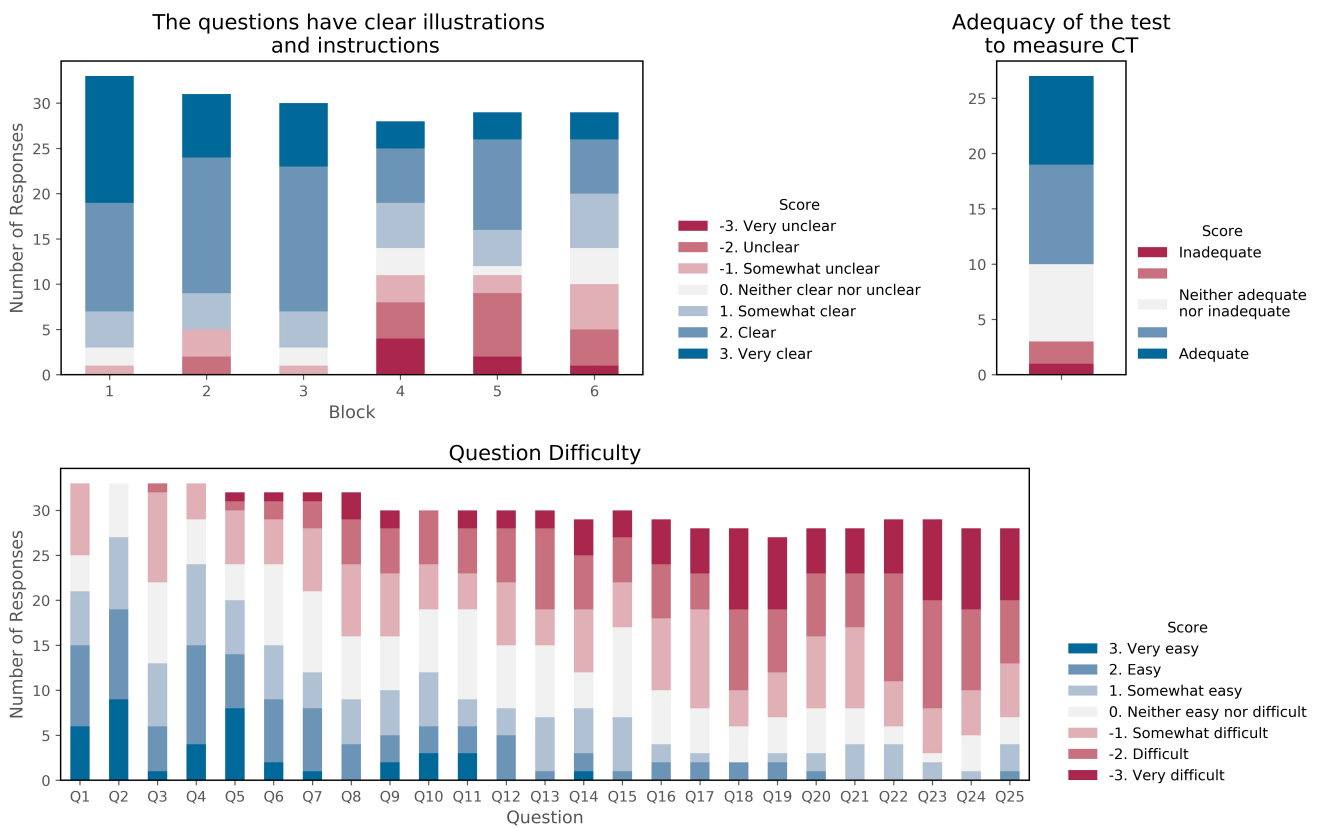

**Figure 3.** Face Validity of the cCTt.

*Construct validity* is established in Fig. 4. The experts were asked to what extent they believed the blocks of questions adequately measure the concepts defined in Table 3. In the worst case, i.e. for the while statements, 76% of the experts agreed that the block of questions adequately measures the targeted concept, with an average agreement of 1.2±0.4% (between "agree" and "totally agree" on a scale of -3 to +3). This seems to point to an adequate construct validity.

*Content validity* was assessed on the basis of whether the experts believed each block of questions was an adequate representation of CT skills. Results were mainly positive with an average approval rating of 1.5±0.3% (on a scale of -3 to +3). In the worst case, a 71% approval was obtained for the conditional statements (see Fig. 4). The experts provided additional insight into their perception of the content validity of the test through the open comments, and in particular in the focus group. The latter, which was unstructured, was primarily focused on what it entails to assess CT considering the fact that i) CT highly multi-dimensional, ii) suffers from a lack of consensus around what CT encompasses, and ii) can be considered both in disciplinary and non-disciplinary contexts. For the experts the cCTt, although it assesses adequately computational concepts, it does not assess CT in all its dimensions. The cCTt, like the CTt and the BCTt before it, mainly focuses on computational concepts and while it includes certain notions of computational practices, it disregards computational perspectives. Indeed, the cCTt does not provide insight into students' thought processes when engaged in CT-related problem solving tasks (Chevalier et al. 2020), their transversal competences (both inter-personal and intrapersonal), and their perception of "of themselves and their relationships with others and the technological world" (Lye and Koh 2014). This is a limitation of most summative assessments (Román-González et al. 2019) which, despite having excellent adequacy for computational concepts and little adequacy for computational practices, are considered inadequate for computational perspectives. Indeed, both the literature and the experts highlighted the importance of combining the cCTt with other tools in a system of assessments to get a comprehensive evaluation of CT interventions (Grover et al. 2015; Román-González et al. 2019). Multiple assessment methods help provide complementary information on the acquisition of CT competences. In particular, computational practices relate to processes and are best assessed through direct observations as in the case of the study by (Chevalier et al. 2020) who looked into the students' thought processes using the Creative Computational Problem Solving model.

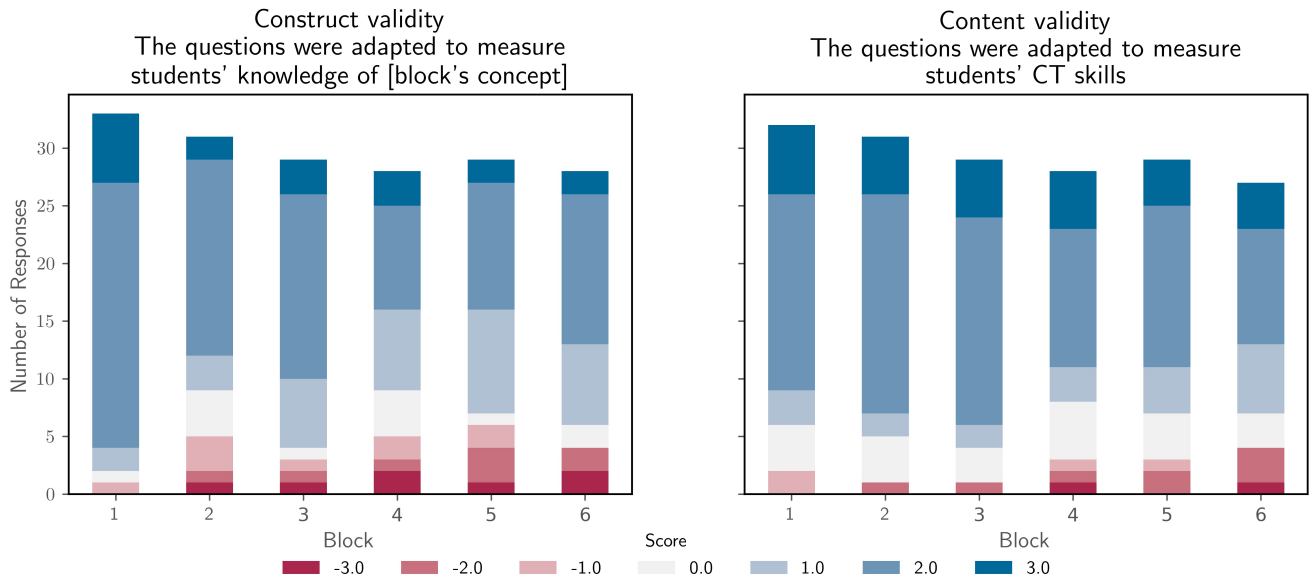

**Figure 4.** Construct and content validity of the cCTt.

*Other suggestions,* both in terms of format and content, emerged from the focus group discussion and open questions of the survey. While we will not detail the format related suggestions that were implemented, we present the remaining content related suggestions that were provided. A primary school educator recommended adding an "I don't know" option so students would not feel pressured into selecting a response. Such an addition also helps distinguish between students who did not have time to answer the question and those who did not know what the correct answer was. Another expert with experience in didactics, pedagogy, research and teacher training also emphasised the importance of establishing error profiles in the selection of responses. Concretely, the idea is that each response should correspond to a type of error, so teachers may also use the test to identify specific learning difficulties and intervene accordingly. Ideally, a teacher would like to know where the students are struggling in order to remedy the situation. Both suggestions were included in the version of the cCTt that was administered to the students (see Table 3 for the error profile attributed to each response).

*3.1.2 Confirmatory Factor Analysis for construct validity*

Based on the test results, a CFA analysis was conducted on the 25 items of the cCTt-25 using the blocks of questions (i.e. CT constructs) as latent factors (see Table 3). This analysis was conducted on three subsets of the data (full, grade 3 and grade 4) to ensure the construct validity was adequate both in general, and for each grade individually. The Kaiser-Meyer-Olkin (KMO) Measure of Sampling Adequacy and Bartlett's test of sphericity were computed to determine whether the data was suitable for structure detection (see Table 7). KMO values were above 0.8 indicating that the sampling is adequate (superior to the 0.5 acceptable limit, Field et al. 2012). Bartlett's test of sphericity is significant with p-values below 0.05, indicating that a factor analysis may be used. Multiple fit indices are then provided for the cCTt-25 model for the different data subsets (see Table 7). The $\chi^2$ statistic is significant in all three cases although this is likely due to the large sample size. That is why the $\chi^2/df$ is considered and is below 3 in all cases indicating an adequate fit. The RMSEA and the SRMR are also below their cutoffs and indicate a good fit. This is also the case for the CFI and TLI which are above the 0.95 cutoff.

Table 8 provides, for each question in the test, its factor and factor loadings. The factor loadings are positive and significant for all questions, with standardised coefficients ranging from 0.540 for question 17 to 0.867 for question 2. Table 9 shows that there are significant positive correlations between the factors themselves. As each factor is related to the performance in a given block of the cCTt, these correlations indicate that students who perform well in one block are likely to perform well in the others.

These results indicate that the cCTt-25 fits well, and that the blocks are representative of the CT concepts they intend to measure.

**Table 7.** Confirmatory Factor Analysis for the cCTt-25, with latent variables and observed variables corresponding to the distribution in Table 3.

| cCTt-25 | Initial conditions (KMO : Kaiser, Meyer, Olkin sampling adequacy) | | Robust model fit indices (CFI : Comparative Fit Index, TLI : Tucker-Lewis Index, RMSEA: Root Mean Square Error of Approximation, SRMR: Standardized Root Mean Square Residual) | | | | | |
|---|---|---|---|---|---|---|---|---|
| | KMO | Bartlett's test of sphericity | $\chi^2$ | $\chi^2/df$ | CFI | TLI | RMSEA | SRMR |
| All | .90 | $\chi^2(300) = 7106, p < .001$ | $\chi^2(260) = 551, p = .000$ | 2.1 | .978 | .974 | .027 | .052 |
| Grade 3 | .88 | $\chi^2(300) = 3101, p < .001$ | $\chi^2(260) = 378, p = .000$ | 1.45 | .977 | .974 | .025 | .061 |
| Grade 4 | .88 | $\chi^2(300) = 3614, p < .001$ | $\chi^2(260) = 403, p = .000$ | 1.6 | .975 | .971 | .027 | .070 |

**Table 8.** cCTt-25 Factor Loadings for CFA on the full dataset.

| Latent Factor | Question | B (factor loading) | Standard error of B | Z-scores | Beta (standardised factor loading) | significance |
|---|---|---|---|---|---|---|
| f1 (sequences) | 1 | 0.699 | 0.053 | 13.154 | 0.699 | *** |
| | 2 | 0.867 | 0.039 | 22.216 | 0.867 | *** |
| | 3 | 0.596 | 0.041 | 14.504 | 0.596 | *** |
| | 4 | 0.694 | 0.033 | 20.775 | 0.694 | *** |
| f2 (simple loops) | 5 | 0.738 | 0.027 | 27.415 | 0.738 | *** |
| | 6 | 0.778 | 0.031 | 24.868 | 0.778 | *** |
| | 7 | 0.635 | 0.030 | 21.125 | 0.635 | *** |
| | 8 | 0.792 | 0.025 | 31.521 | 0.792 | *** |
| f3 (complex loops) | 9 | 0.608 | 0.032 | 19.013 | 0.608 | *** |
| | 10 | 0.715 | 0.025 | 29.122 | 0.715 | *** |
| | 11 | 0.771 | 0.021 | 36.155 | 0.771 | *** |
| | 12 | 0.769 | 0.021 | 36.626 | 0.769 | *** |
| | 13 | 0.844 | 0.019 | 45.050 | 0.844 | *** |
| | 14 | 0.612 | 0.027 | 22.423 | 0.612 | *** |
| | 15 | 0.768 | 0.022 | 34.956 | 0.768 | *** |
| f4 (conditionals) | 16 | 0.721 | 0.030 | 24.231 | 0.721 | *** |
| | 17 | 0.540 | 0.047 | 11.494 | 0.540 | *** |
| | 18 | 0.688 | 0.029 | 23.688 | 0.688 | *** |
| | 19 | 0.652 | 0.030 | 21.765 | 0.652 | *** |
| f5 (while statements) | 20 | 0.626 | 0.032 | 19.645 | 0.626 | *** |
| | 21 | 0.765 | 0.026 | 29.423 | 0.765 | *** |
| | 22 | 0.576 | 0.034 | 16.903 | 0.576 | *** |
| | 23 | 0.540 | 0.034 | 15.862 | 0.540 | *** |
| f6 (combinations) | 24 | 0.547 | 0.049 | 11.102 | 0.547 | *** |
| | 25 | 0.836 | 0.050 | 16.786 | 0.836 | *** |

**Table 9.** cCTt-25 Latent Factor Correlations for CFA on the full dataset.

| Factor 1 | f1 | f1 | f1 | f1 | f1 | f2 | f2 | f2 | f2 | f3 | f3 | f3 | f4 | f4 | f5 |
|---|---|---|---|---|---|---|---|---|---|---|---|---|---|---|---|
| Factor 2 | f2 | f3 | f4 | f5 | f6 | f3 | f4 | f5 | f6 | f4 | f5 | f6 | f5 | f6 | f6 |
| Correlation | .686 | .663 | .559 | .489 | .367 | .758 | .552 | .559 | .426 | .607 | .583 | .510 | .803 | .671 | .808 |
| Significance | *** | *** | *** | *** | *** | *** | *** | *** | *** | *** | *** | *** | *** | *** | *** |

## 3.2    RQ2 - Reliability of the cCTt

Reliability of the cCTt was established through two means which are considered complimentary: Classical Test Theory (see section 3.2.1) and Item Response Theory (see section 3.2.2). Both approaches appear to indicate that the test has a wide range of question difficulties and adequately discriminates between students. The IRT contributes to this by showing that the test is best suited to discriminate between students with low and medium abilities, while classical test theory adds that the cCTt has a good level of internal consistency and exhibits no clear ceiling or flooring effects, with students in grade 4 scoring significantly better than students in grade 3.

*3.2.1 Reliability through classical test theory and item analysis* The distribution of scores obtained by the students is shown in Fig. 5. Students score on average 14.14±5.22 out of a total of 25, with no evident ceiling effect, and 8% of students scoring below chance (i.e. < 25/4). Considering the minimum effect size that can be considered in the study to achieve a power of 0.8 (Cohen's $D = .14$), significant differences are observed between grades ($p < 0.001$, +2.9pt in grade 4, Cohen's $D = 0.57$), and gender ($p = 0.013$, +0.6pt for boys, Cohen's $D = 0.15$). Although the effect size for gender is small, this brings up the question of when (and *why*) gender differences arise in STEM and disciplines related to Computer Science (CS), and the need to effectively addressed them in ongoing CS and CT curricular reforms (El-Hamamsy et al. 2021b) . Provided the small effect size for gender, we only distinguish between grades in the rest of the classical test theory analysis.

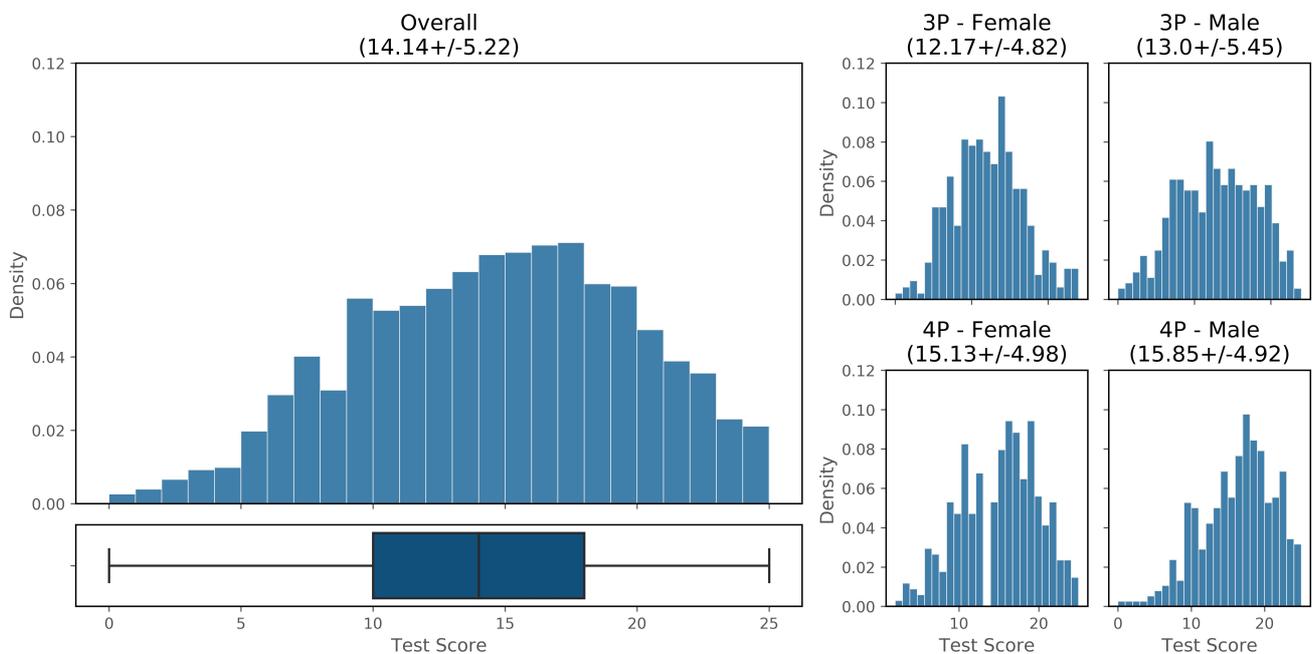

**Figure 5.** Score distribution per grade and gender differences.

Fig. 6 shows the proportion of correct responses for students in grade 3, grade 4 and for the full sample. The questions appear to be increasingly difficult with certain earlier questions being too easy (difficulty index > 0.85 for questions 1 and 2 for all, and question 6 for grade 4), and later questions being too hard (difficulty index < 0.25 for question 17 and 24 for all, and question 25 for grade 3). Students in grade 4 appear to score consistently better than students in grade 3 over all questions. Looking at the point-biserial correlation, which is the difference between the high scorers and the low scorers of the sample population, all questions score above 0.2 which is considered acceptable.

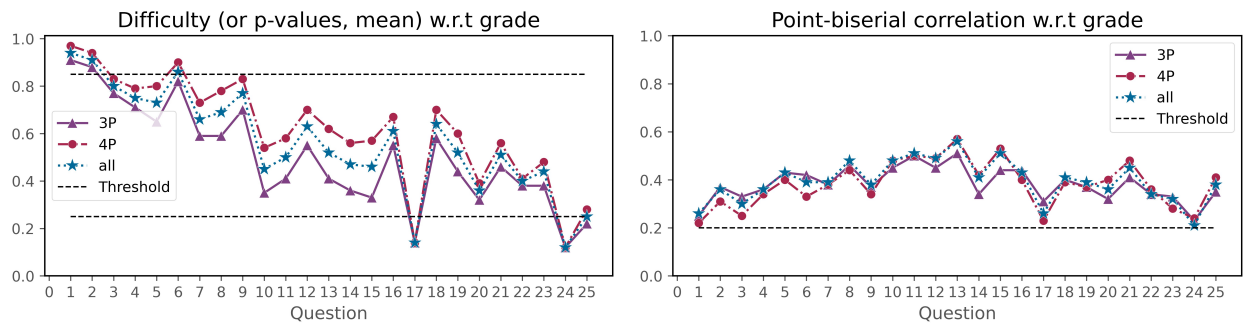

**Figure 6.** Question difficulty and point-biserial correlation distribution per question. Items with a difficulty below .25 are considered too hard while items with a difficulty above .85 are considered too easy. A point-biserial correlation between .2 and .3 is considered good, and above .3 excellent.

Finally, we consider the reliability of the cCTt with respect to Cronbach's $\alpha$ measure of internal consistency. The overall test reliability is high (Taherdoost 2016) with an alpha of .85 for the full sample and .84 for grade 3 and grade 4 individually. The drop alpha was computed for each question in the test to determine whether removing a given question would improve the internal consistency of the test. The results in Table 10 show that internal consistency does not improve when removing questions. When considering the reliability of the individual blocks of questions, whether for the full sample of each grade, the reliability is high for blocks 1-3 and 5 ($0.7 < \alpha < 0.9$), and moderate for 4 and 6 ($0.5 < \alpha < 0.7$), according to the thresholds of Hinton et al. (2014) (referenced by Taherdoost 2016), yielding acceptable levels of internal consistency for psychological assessments (Relkin et al. 2020).

*3.2.2 Reliability through item response theory* Item Response Theory complements the results of the classical test theory analysis by considering the relationship between a) the probability that a student answers a question correctly and b) an underlying latent ability. Two models were tested: i) the one-parameter model (called 1-Parameter Logistic, or 1-PL) which considers that only difficulty varies between items, and ii) the two-parameter model (called 2-Parameter Logistic, or 2-PL), which considers that both difficulty and discrimination varies across items. Table 11 shows the comparison between the two models. Here, the test is significant ($p < 0.001$) indicating that one model offers a better fit. As lower AIC and BIC scores are obtained with the 2-PL model, this indicates that the 2-PL model is to be preferred.

The IRT analysis results with the 2-PL model are shown in Fig. 7, 8 and 9. The Item Characteristic Curves (or ICC) show the probability of selecting the correct response for a given question of the cCTt with respect to the students' ability. In the Fig., easier questions are those with a curve which begins to increase already at lower abilities. Additionally, questions which better discriminate between students with high and low ability are those with a steeper slope. The objective is to have a test that is composed of questions with varying degrees of difficulty, thus covering a wide spectrum of abilities, and which discriminate well (steep slopes). The ICCs in Fig. 7 show curves at varying levels of ability, with easy and medium questions discriminating better than harder ones. Indeed the average test difficulty is equal to 0.032±1.2 on the logit scale, with the easiest question yielding a difficulty index of −2.4 and the hardest of 2.8 (see Table 12). The Item Information Curves (IIC) in Fig. 8 offer complementary information by indicating the amount of information that each question provides for a given ability. This means that more precise items at a given ability level are higher on the IIC scale. As higher peaks appear in the low to medium ability range, the test would appear to be better suited to discriminate between students at that level. This is confirmed by the Test Information Function (TIF, sum of the IICs, Hambleton and Jones 1993) in Fig. 9 which is centred around 0 (medium ability) and slightly higher on the left than on the right, thus providing more information about students with lower abilities than with higher abilities.

**Table 10.** Classical test theory item analysis. Items with a difficulty below .25 are considered too hard while items with a difficulty above .85 are considered too easy. A point-biserial correlation between .2 and .3 is considered good, and above .3 excellent. Q : question, 3P : grade 3, 4P : grade 4.

| Block | Q | Block alpha | | | 1. Difficulty (or p-values, mean) | | | 2. Standard Deviation | | | 3. Point-biserial correlation (or item discrimination, item-total correlation) | | | Individual drop alpha | | |
|---|---|---|---|---|---|---|---|---|---|---|---|---|---|---|---|---|
| | | all | 3P | 4P | all | 3P | 4P | all | 3P | 4P | all | 3P | 4P | all | 3P | 4P |
| 1 | 1 | | | | .94 | .91 | .97 | .23 | .28 | .17 | .26 | .25 | .22 | .85 | .84 | .84 |
| | 2 | | | | .91 | .88 | .94 | .28 | .33 | .24 | .36 | .37 | .31 | .85 | .84 | .84 |
| | 3 | .76 | .75 | .79 | .8 | .77 | .83 | .4 | .42 | .38 | .3 | .33 | .25 | .85 | .84 | .84 |
| | 4 | | | | .75 | .71 | .79 | .43 | .46 | .41 | .36 | .36 | .34 | .85 | .84 | .84 |
| 2 | 5 | | | | .73 | .65 | .8 | .44 | .48 | .4 | .43 | .43 | .4 | .84 | .84 | .84 |
| | 6 | | | | .86 | .82 | .9 | .34 | .38 | .3 | .39 | .42 | .33 | .85 | .84 | .84 |
| | 7 | .78 | .77 | .77 | .66 | .59 | .73 | .47 | .49 | .44 | .39 | .38 | .38 | .85 | .84 | .84 |
| | 8 | | | | .69 | .59 | .78 | .46 | .49 | .41 | .48 | .46 | .44 | .84 | .83 | .84 |
| 3 | 9 | | | | .77 | .7 | .83 | .42 | .46 | .38 | .38 | .37 | .34 | .85 | .84 | .84 |
| | 10 | | | | .45 | .35 | .54 | .5 | .48 | .5 | .48 | .45 | .48 | .84 | .83 | .83 |
| | 11 | | | | .5 | .41 | .58 | .5 | .49 | .49 | .51 | .5 | .5 | .84 | .83 | .83 |
| | 12 | .83 | .81 | .83 | .63 | .55 | .7 | .48 | .5 | .46 | .49 | .45 | .49 | .84 | .83 | .83 |
| | 13 | | | | .52 | .41 | .62 | .5 | .49 | .49 | .56 | .51 | .57 | .84 | .83 | .83 |
| | 14 | | | | .47 | .36 | .56 | .5 | .48 | .5 | .41 | .34 | .42 | .85 | .84 | .84 |
| | 15 | | | | .46 | .33 | .57 | .5 | .47 | .5 | .51 | .44 | .53 | .84 | .83 | .83 |
| 4 | 16 | | | | .61 | .55 | .67 | .49 | .5 | .47 | .43 | .44 | .4 | .84 | .83 | .84 |
| | 17 | | | | .14 | .14 | .14 | .34 | .34 | .35 | .26 | .31 | .23 | .85 | .84 | .84 |
| | 18 | .67 | .7 | .62 | .64 | .58 | .7 | .48 | .49 | .46 | .41 | .4 | .39 | .85 | .84 | .84 |
| | 19 | | | | .52 | .44 | .6 | .5 | .5 | .49 | .39 | .37 | .37 | .85 | .84 | .84 |
| 5 | 20 | | | | .36 | .32 | .39 | .48 | .47 | .49 | .36 | .32 | .4 | .85 | .84 | .84 |
| | 21 | | | | .51 | .46 | .56 | .5 | .5 | .5 | .45 | .41 | .48 | .84 | .84 | .83 |
| | 22 | .72 | .7 | .73 | .4 | .38 | .41 | .49 | .49 | .49 | .34 | .34 | .36 | .85 | .84 | .84 |
| | 23 | | | | .44 | .38 | .48 | .5 | .49 | .5 | .32 | .33 | .28 | .85 | .84 | .84 |
| 6 | 24 | .57 | .58 | .57 | .12 | .12 | .12 | .32 | .33 | .33 | .21 | .23 | .24 | .85 | .84 | .84 |
| | 25 | | | | .25 | .22 | .28 | .43 | .41 | .45 | .38 | .35 | .41 | .85 | .84 | .84 |
| Overall | | .85 | .85 | .83 | .59 | .53 | .65 | .49 | .50 | .48 | - | - | - | - | - | - |

**Table 11.** Comparison of the 1-PL and 2-PL model.

| Model | AIC | BIC | log likelihood | LRT | df | p.value |
|---|---|---|---|---|---|---|
| 1-PL | 39156.40 | 39294.87 | −19552.20 | | | |
| 2-PL | 38904.32 | 39170.61 | −19402.16 | 300.08 | 24 | < 0.001 |

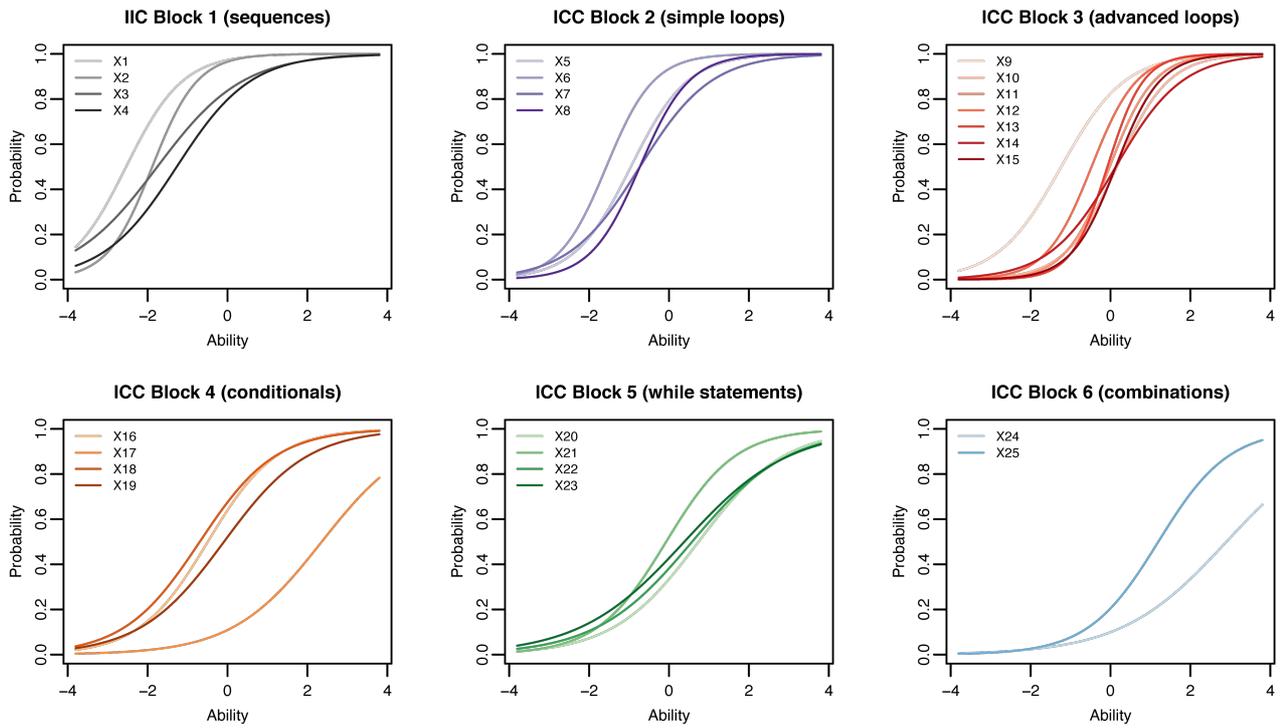

**Figure 7.** 2-PL Model IRT Item Characteristic Curves (ICC).

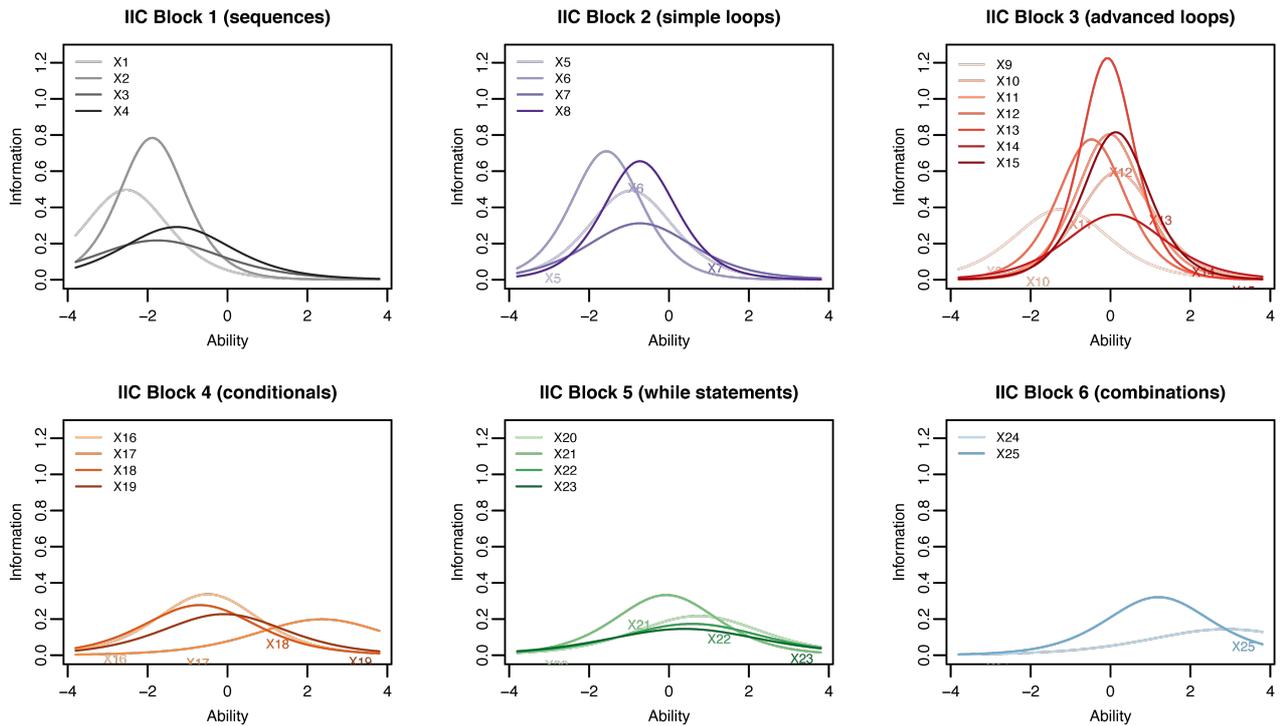

**Figure 8.** 2-PL Model IRT Item Information Curves (IIC).

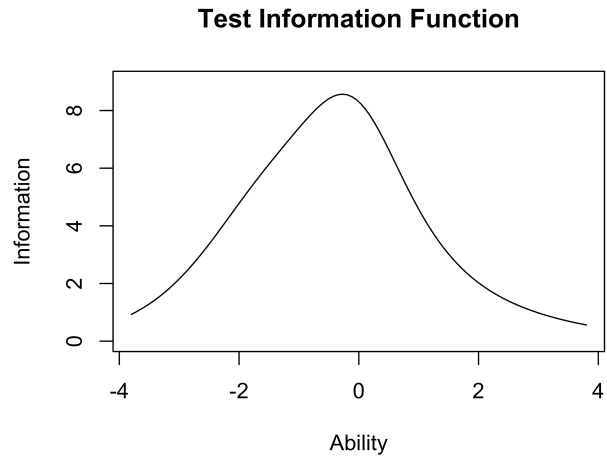

**Figure 9.** 2-PL Model IRT Test Information Function (TIF).

**Table 12.** IRT Item difficulty and discrimination indices.

| Question | IRT Item difficulty index | IRT Item discrimination index |
|---|---|---|
| 1 | $-2.40$ | 1.18 |
| 2 | $-1.67$ | 1.70 |
| 3 | $-1.42$ | 1.03 |
| 4 | $-1.01$ | 1.07 |
| 5 | $-0.63$ | 1.28 |
| 6 | $-1.28$ | 1.75 |
| 7 | $-0.43$ | 1.03 |
| 8 | $-0.37$ | 1.41 |
| 9 | $-0.92$ | 1.13 |
| 10 | 0.57 | 1.43 |
| 11 | 0.30 | 1.68 |
| 12 | $-0.19$ | 1.44 |
| 13 | 0.32 | 1.78 |
| 14 | 0.69 | 0.95 |
| 15 | 0.67 | 1.42 |
| 16 | $-0.23$ | 1.18 |
| 17 | 2.07 | 1.08 |
| 18 | $-0.37$ | 1.01 |
| 19 | 0.29 | 0.93 |
| 20 | 1.04 | 0.80 |
| 21 | 0.16 | 1.03 |
| 22 | 0.66 | 0.83 |
| 23 | 0.65 | 0.83 |
| 24 | 2.80 | 0.79 |
| 25 | 1.47 | 1.06 |
| $\mu \pm \sigma$ | $0.032 \pm 1.2$ | $1.2 \pm 0.31$ |

## 3.3 RQ3 - Shortening the cCTt through Confirmatory Factor Analysis

Administration times lasted on average 35 min for students in grade 3 and 30 min for students in grade 4, according to an estimation with 453 grade 3 students and 680 grade 4 students. Teachers reported having taken a an average of 65±35 min (minimum 25 min, median of 50 min, and a maximum of 4 sessions) to administer the test, including explanations, examples provided to students and breaks. Provided the expert evaluation where 41% of experts considered that the test was too long, as well as the teachers' feedback following the test administration, in addition to the reported duration of administration, it is important to consider how the test may be shortened. Since the results of the Classical Test Theory analysis (see section 3.2.1) showed that the problematic questions were the same for students in grade 3 and grade 4 alike (despite students in grade 3 scoring less), we believe that there is not enough difference to justify having an ad-hoc test for each grade. That is why we chose to apply the same shortening procedure, independently of the students' grade, to propose several shorter versions of the cCTt which we validate on the full dataset. The chosen approach is Confirmatory Factor Analysis, more specifically using the provided modification indices (i.e. the amount by which the $\chi^2$ statistic would improve) to identify questions which are highly correlated and thus redundant, or load on other factors (and thus reduce the model's fit). Therefore, taking into consideration that we would like to have blocks which are somewhat balanced in length, we iteratively removed questions as described in Table 13. The resulting variants of the test are presented in Table 14:

- The cCTt-25, the longest version of the test which meets all the traditional thresholds for the model fit statistics.
- The cCTt-17, which covers the same breadth of constructs of the cCTt-25 but lacks a certain number of redundancies over the different blocks of questions.
- The cCTt-15, the fastest and most focused test, which does not include the final block of questions that evaluate the combination of concepts and were the most difficult for students.

**Table 13.** Iterative procedure to shorten the cCTt based on CFA.

| Model | Successive editions to the cCTt-25 | Model fit statistics $\chi^2$ | $\chi^2/df$ | CFI | TLI | RMSEA | SRMR |
|---|---|---|---|---|---|---|---|
| cCTt-17 | 1. Removing Q17: correlates with Q24 & has a high difficulty | $\chi^2(237) = 464, p = .000$ | 1.96 | .982 | .979 | .025 | .047 |
| | 2. Removing Q22: correlates with Q2 and also loads on factor 4 | $\chi^2(215) = 394, p = .000$ | 1.83 | .985 | .983 | .023 | .045 |
| | 3. Removing Q9: also loads on factors 1 & 2 | $\chi^2(194) = 336, p = .000$ | 1.73 | .988 | .986 | .022 | .044 |
| | 4. Removing Q10: correlates with Q4 & loads on factors 1 & 2 | $\chi^2(174) = 294, p = .000$ | 1.69 | .989 | .986 | .021 | .043 |
| | 5. Removing Q2: correlates with Q4, & is too easy | $\chi^2(155) = 260, p = .000$ | 1.68 | .989 | .987 | .021 | .041 |
| | 6. Removing Q4: loads on factors 4, 5 and 6 | $\chi^2(137) = 228, p = .000$ | 1.66 | .990 | .988 | .021 | .040 |
| | 7. Removing Q8: also loads on factor 3 | $\chi^2(120) = 201, p = .000$ | 1.68 | .991 | .988 | .021 | .040 |
| | 8. Removing Q14 (last remaining canvas question) | $\chi^2(104) = 176, p = .000$ | 1.69 | .991 | .988 | .021 | .040 |
| cCTt-15 | 9-10. Removing Q24 & Q25 (block 6, the combination of constructs) | $\chi^2(80) = 137, p = .000$ | 1.71 | .992 | .989 | .022 | .038 |

**Table 14.** Questions in the variants of the cCTt.

| | 1 | 2 | 3 | 4 | 5 | 6 | 7 | 8 | 9 | 10 | 11 | 12 | 13 | 14 | 15 | 16 | 17 | 18 | 19 | 20 | 21 | 22 | 23 | 24 | 25 |
|---|---|---|---|---|---|---|---|---|---|---|---|---|---|---|---|---|---|---|---|---|---|---|---|---|---|
| cCTt-25 | x | x | x | x | x | x | x | x | x | x | x | x | x | x | x | x | x | x | x | x | x | x | x | x | x |
| cCTt-17 | x | | x | | x | x | x | | | | x | x | x | | x | x | | x | x | x | x | | x | x | x |
| cCTt-15 | x | | x | | x | x | x | | | | x | x | x | | x | x | | x | x | x | x | | x | | |

## 4   Discussion and Conclusion

With the introduction of CT in curricula worldwide, there is a pressing need to have validated and reliable instruments to assess Computational Thinking throughout mandatory schooling. It is not surprising to see that CT assessment is one of the most prominent topics in CT research (Tikva and Tambouris 2021), with researchers working to develop instruments from kindergarten (Relkin et al. 2021), through primary school (Relkin et al. 2020; Zapata-Cáceres et al. 2020) and up to middle school (Román-González et al. 2017, 2019). However, and to the best of our knowledge, none have proven to be both valid and reliable measurements of CT in upper primary school. Therefore, building up on the CT test (Román-González et al. 2017, 2019) designed for middle school, and the subsequent BCT test which adapted the CT test and validated it for use in lower primary school, we developed the cCT test to pally the lack of validated instruments in upper primary school.

The cCT test is an unplugged CT assessment adapted from the BCT test in terms of format and content, to be administered to students in the 7-9 age range regardless of their prior coding experience. The test is composed of 25 multiple choice questions of increasing difficulty and addressing notions of sequences, loops, conditionals and while statements. To assess the tests' validity (face, construct, and content validity) we conducted an expert evaluation with 37 participants. The survey results and focus group indicate that the test has good face, content and construct validity. Then, to validate the psychometric properties of the cCTt (in terms of validity and reliability), the test was administered to 1519 students from 77 classes in grades 3 and 4 (ages 7-9) enrolled in 7 schools of the same region. The analysis of the results involved three stages and the outcomes can be summarised as follows. In the first stage, Confirmatory Factor Analysis confirmed the construct validity of the different blocks of the test. In the second stage, the results from Classical Test Theory showed that there was no evident ceiling or flooring effect, with scores distributed around 14/25, although students in grade 3 score significantly lower than those in grade 4. The Classical Test Theory analysis also indicated adequate reliability with good internal consistency (Cronbach's $\alpha$ = 0.85), levels of discrimination (Point biserial correlations > 0.2) and a wide range of question difficulties (proportion of correct responses). In the final stage, the Item Response Theory analysis supported these findings and further indicated that the test was better suited at evaluating and discriminating between students with low and medium abilities. In addition to the psychometric analyses, and to pally the limitations posed by administration time that were brought up both by teachers and experts, several shortened versions of the cCT test are proposed (cCTt-17 and cCTt-15), having been established through an iterative shortening procedure using Confirmatory Factor Analysis.

While the test has adequate face, content, and construct validity, as determined through expert validation, and good psychometric properties, there are two main limitations. Firstly, to achieve a more exhaustive measurement of CT is important to consider combining assessments such as the cCTt with other forms of assessments (Grover et al. 2015; Román-González et al. 2019), thus improving the content validity of the overall assessment scheme. This is because the cCTt, as a paper-based test, does not measure all aspects of CT. The cCTt specifically focuses on computational concepts and practices while lacking computational perspectives (Román-González et al. 2019), and more generally competences (Lye and Koh 2014), an identified limitation of many summative CT assessments. Secondly, criterion validity needs to be established with respect to a "gold standard". Three main approaches exist. Typically, researchers make a comparison with other assessment methods (convergent validity, Relkin et al. 2020, 2021; Román-González et al. 2017). More classically however, criterion validity is established through

   i)    determining the test's predictive validity (i.e. does the test predict something that it should predict, such as academic performance and coding achievement, as done in Román-González et al. 2018)
   ii)   determining its concurrent validity (can the test distinguish between two populations that are distinct, e.g. can we distinguish between students who partake in CT related activities and those who don't?).

Provided that validation is a multi-step process which requires "collect[ing] multiple sources of evidence to support the proposed interpretation and use of assessment result[s]" and "multiple methodologies, sources of data, and types of analysis" (Gane et al. 2021), future work should continue to validate the instrument in consideration of the test's content and criterion validity, for both the target and older age groups (namely grades 5 and 6).

To conclude, the cCTt appears to be a valid and reliable instrument for CT assessment in grades 3 and 4 (students aged 7-9), which should be combined with other assessments to include computational perspectives and practices. The test is easy to administer and score on a large scale. With this work we therefore extend the portfolio of CT assessments designed for use by researchers and teachers in formal education. In particular, the BCTt, cCTt and CTt now jointly cover the range needed for CT assessment throughout primary and secondary school, by means of unplugged tests. It would be desirable, however, to extend the study by applying the cCTt to other populations, and to specifically study the age limits for using one test or the other.

**Notes**

1. A competence refers to "the proven ability to use knowledge, skills and personal, social and/or methodological abilities, in work or study situations and in professional and personal development" (European Union 2006).

2. Computer Science Unplugged activities (Bell and Vahrenhold 2018) are defined as activities which develop core CS competences without using screens. These activities have numerous advantages including i) making use of embodied cognition, ii) being adapted to a wide range of learners by reducing the cognitive load pertaining to use of technological devices and programming artefacts (Romero et al. 2018), iii) saving time pertaining to the mastery of specific tools and programming interfaces (Webb et al. 2017), and iv) not requiring any specific technological devices. Owing to these benefits, CS unplugged activities have gained in popularity over the past two decades, with the term "unplugged" now referring to a type of pedagogy (Bell and Vahrenhold 2018) with a wide range of applications in outreach initiatives and more recently teacher training and classrooms to support formal curricula. Once example which can be cited is the case of Switzerland where the lower primary school (grades 1-4) CS curricular reform is heavily focused on unplugged content which teachers favour and employ more largely than their plugged counterparts (El-Hamamsy et al. 2021a) in big part due to a reticence towards the presence of screens in classrooms (Negrini 2020). While there is often a debate around the relevance of such types of activities to develop CT competences, there is an increasing amount of research being done on the topic. While certain small scale studies were undertaken and showed that CS Unplugged activities could be as effective as traditional approaches (Thies and Vahrenhold 2013, 2016; Hermans and Aivaloglou 2017), more and more large scale studies at the level of primary school show the benefits of CS Unplugged activities compared to traditional approaches for learning (Brackmann et al. 2017; del Olmo-Muñoz et al. 2020; Sun et al. 2021; Zhan et al. 2022; Kirc¸ali and Ozdener¨ 2022), in addition to the benefits in terms of motivation and gender issues (del Olmo-Muñoz et al. 2020), engagement (Zhan et al. 2022), and self-efficacy (Hermans and Aivaloglou 2017), thus contributing to the promotion of CS for all and the development of CT competencies (Huang and Looi 2021).

3. CT skills here refers to the definition of skills provided by the European Union (2006) as "the ability to apply knowledge and use know-how to complete tasks and solve problems". Measuring CT skills thus implies assessing the outcome of the application of the knowledge of the underlying CT concepts, without looking into the processes involved. Such assessments therefore do not evaluate the full range of CT competences involved in the resolution of CT-tasks.

4. See the section of the Encyclopedia Britannica on Psychometric Theories here https://www.britannica.com/science/ human-intelligence-psychology/Psychometric-theories.

5. Please note that experts considered this question from a content perspective, i.e. whether the test grasps all facets of CT, despite the use of the word "evaluation".

6. See https://www.auto-multiple-choice.net/index.en for the link to the Auto-Multiple-Choice software.

**Data availability**

The data used for the study is available on Zenodo at https://doi.org/10.5281/zenodo.5865572.

**Acknowledgements**

This document is the result of the research project funded by the Swiss National Science Foundation's National Center of Competence in Research (NCCR) Robotics and the Madrid Regional Government, through the project e- Madrid-CM (P2018/TCS-4307) which is also co-financed by the Structural Funds (FSE and FEDER). We would like to thank all those who were implicated in the expert evaluation, test administration in the schools, as well as the teachers and students who participated in the validation process.

# A  cCTt-25 question set

**Figure 10.** Questions of the block 1 (sequences) of the cCTt.

**Figure 11.** Questions of the block 2 (simple loops) of the cCTt.

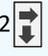

**Figure 12.** Questions of the block 3 (complex loops) of the cCTt.

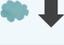

**Figure 13.** Questions of the block 4 (conditionals) of the cCTt.

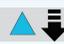

**Figure 14.** Questions of the block 5 (while statements) of the cCTt.

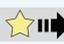

**Figure 15.** Questions of the block 6 (combination of concepts) of the cCTt.

# B     cCTt Example questions

**Figure 16.** Example questions of the cCTt.

## C      Factor Loadings for the Confirmatory Factor Analysis of the cCTt variants

### C.1    cCTt-17

**Table 15.** cCTt-17 Factor Loadings for CFA on the full dataset.

| Latent Factor | Question | B (factor loading) | Standard error of B | Z-scores | Beta (standardised factor loading) | significance |
|---|---|---|---|---|---|---|
| f1 | 1 | 0.694 | 0.068 | 10.281 | 0.694 | *** |
| (sequences) | 3 | 0.612 | 0.051 | 11.993 | 0.612 | *** |
|  | 5 | 0.759 | 0.030 | 25.253 | 0.759 | *** |
| f2 | 6 | 0.805 | 0.034 | 24.021 | 0.805 | *** |
| (simple loops) | 7 | 0.640 | 0.033 | 19.133 | 0.640 | *** |
|  | 11 | 0.784 | 0.023 | 34.546 | 0.784 | *** |
| f3 | 12 | 0.807 | 0.021 | 38.091 | 0.807 | *** |
| (complex loops) | 13 | 0.857 | 0.021 | 41.788 | 0.857 | *** |
|  | 15 | 0.786 | 0.023 | 33.875 | 0.786 | *** |
|  | 16 | 0.721 | 0.031 | 23.437 | 0.721 | *** |
| f4 | 18 | 0.702 | 0.030 | 23.446 | 0.702 | *** |
| (conditionals) | 19 | 0.668 | 0.031 | 21.382 | 0.668 | *** |
|  | 20 | 0.644 | 0.033 | 19.690 | 0.644 | *** |
| f5 | 21 | 0.798 | 0.028 | 28.793 | 0.798 | *** |
| (while statements) | 23 | 0.585 | 0.034 | 17.341 | 0.585 | *** |
| f6 | 24 | 0.520 | 0.051 | 10.279 | 0.520 | *** |
| (combinations) | 25 | 0.879 | 0.056 | 15.683 | 0.879 | *** |

**Table 16.** cCTt-17 Latent Factor Correlations for CFA on the full dataset.

| Factor 1 | f1 | f1 | f1 | f1 | f1 | f2 | f2 | f2 | f2 | f3 | f3 | f3 | f4 | f4 | f5 |
| --- | --- | --- | --- | --- | --- | --- | --- | --- | --- | --- | --- | --- | --- | --- | --- |
| Factor 2 | f2 | f3 | f4 | f5 | f6 | f3 | f4 | f5 | f6 | f4 | f5 | f6 | f5 | f6 | f6 |
| Correlation | .792 | .615 | .528 | .365 | .300 | .676 | .554 | .517 | .407 | .561 | .551 | .474 | .730 | .569 | .756 |
| Significance | *** | *** | *** | *** | *** | *** | *** | *** | *** | *** | *** | *** | *** | *** | *** |

## C.2 cCTt-15

**Table 17.** cCTt-15 Factor Loadings for CFA on the full dataset.

| Latent Factor | Question | B (factor loading) | Standard error of B | Z-scores | Beta (standardised factor loading) | significance |
|---|---|---|---|---|---|---|
| f1 (sequences) | 1 | 0.697 | 0.067 | 1.350 | 0.697 | *** |
|  | 3 | 0.609 | 0.051 | 11.909 | 0.609 | *** |
|  | 5 | 0.759 | 0.030 | 25.492 | 0.759 | *** |
| f2 (simple loops) | 6 | 0.808 | 0.034 | 24.118 | 0.808 | *** |
|  | 7 | 0.636 | 0.033 | 19.141 | 0.636 | *** |
|  | 11 | 0.788 | 0.023 | 34.896 | 0.788 | *** |
| f3 (complex loops) | 12 | 0.809 | 0.021 | 38.214 | 0.809 | *** |
|  | 13 | 0.856 | 0.021 | 41.606 | 0.856 | *** |
|  | 15 | 0.780 | 0.024 | 33.123 | 0.780 | *** |
|  | 16 | 0.721 | 0.031 | 22.954 | 0.721 | *** |
| f4 (conditionals) | 18 | 0.716 | 0.030 | 23.730 | 0.716 | *** |
|  | 19 | 0.654 | 0.032 | 20.496 | 0.654 | *** |
|  | 20 | 0.643 | 0.035 | 18.635 | 0.643 | *** |
| f5 (while statements) | 21 | 0.788 | 0.030 | 26.665 | 0.788 | *** |
|  | 23 | 0.596 | 0.035 | 16.989 | 0.596 | *** |

**Table 18.** cCTt-15 Latent Factor Correlations for CFA on the full dataset.

| Factor 1 | f1 | f1 | f1 | f1 | f2 | f2 | f2 | f3 | f3 | f4 |
|---|---|---|---|---|---|---|---|---|---|---|
| Factor 2 | f2 | f3 | f4 | f5 | f3 | f4 | f5 | f4 | f5 | f5 |
| Correlation | .792 | .616 | .530 | .366 | .676 | .553 | .519 | .561 | .552 | .731 |
| Significance | *** | *** | *** | *** | *** | *** | *** | *** | *** | *** |

**Author Bios**

*Laila El-Hamamsy* is a PhD candidate at the Ecole Polytechnique Fédérale de Lausanne (EPFL) with the MOBOTS Group and LEARN, the Center for Learning Sciences. Her research focuses the introduction of Computer Science and Educational Robotics into teacher practices, within the context of a digital education curricular reform and the associated teacher professional development program. Prior to that, she received an MSc. in Robotics and a BSc. in Micro-engineering from EPFL.

*María Zapata-Cáceres* received the Architect BSc Degree and MEng at the Universidad Politecnica de Madrid (UPM) in 2002, M.S. Degrees in Virtual Environments (CSA) in 2002, and Videogames Design and Production (UEM) in 2005, and Computer Science BSc Degree at Universidad Nacional de Educacion a Distancia (UNED) in 2018. She is currently a researcher and visiting professor in the Games Design and Development BSc Degree at the Computer Science Department at the Universidad Rey Juan Carlos (URJC) in Madrid. Her main area of research includes video games as learning instruments for computer science. She has more than 15 years of professional experience as an entrepreneur with activities related to 3D design, video games, technology, and teaching.

*Estefanía Martín Barroso* received the Computer Science BSc Degree and MEng Degree in 2002 and 2004. She obtained her PhD degree in Computer Science and Telecommunications in 2008 at Universidad Autonoma de Madrid (UAM). Her thesis was focused on mobile adaptive learning systems for collaborative contexts.From 2003 to 2008, she was an Assistant Professor at UAM and Universidad Rey Juan Carlos (URJC). Currently, she is an associate professor in the Computer Science Department at the URJC. She leads Blue Thinking project, an application that allows the person with ASD to learn programming; DEDOS project, which provides authoring tools for creating educational activities on multiple devices, and ClipIt, a video-based social network platform developed in the EU project JuxtaLearn. Her research interests include learning systems, HCI and disabilities.

*Francesco Mondada* is a professor at the Ecole Polytechnique Fédérale de Lausanne (EPFL), Switzerland and director´ of the Center for Learning Sciences at EPFL. After a master and a PhD received at EPFL, he led the design of many miniature mobile robots, commercialised and used worldwide in thousands of schools and universities. He co-founded several companies selling these robots or other educational tools. He is author of more than a hundred publications in the field of robot design. He received several awards, including the Swiss Latsis University prize, as best young researcher at EPFL and the Credit Suisse Award for Best Teaching as best teacher at EPFL.

*Jessica Dehler Zufferey* earned her PhD from University of Tubingen in the field of computer-supported collaborative learning (CSCL) with a focus on knowledge awareness. She held postdoctoral positions at University of Fribourg in university didactics, working on gender neutral teaching, and at the Ecole Polytechnique Fédérale de Lausanne (EPFL) in the Computer-Human-Interaction in Learning and Instruction CHILI lab, working on technology-enhanced vocational training. After 4 years in a RD role in the educational technology industry, she is now executive director of the EPFL Center for Learning Sciences.

*Barbara Bruno* is a post-doctoral researcher at the Ecole Polytechnique Fédérale de Lausanne (EPFL), in Lausanne, Switzerland, in the CHILI lab. She is a co-founder of the start-up company Teseo, Italy. Barbara received the M.Sc. and the Ph.D. in Robotics from the University of Genoa in 2011 and 2015, respectively. She is part of the NCCR Robotics organisation and currently involved in the EU ITN ANIMATAS. In 2017-2019 she was Technical Manager of the H2020 project CARESSES. She has published more than 45 articles in international journals and peer-reviewed international conferences. Her research interests are in Social Robotics and Human–Robot Interaction and Cooperation.